\documentclass[10pt,letterpaper,twocolumn]{article} %% two column, final layout

\usepackage{ol2}
\usepackage[draft]{hyperref}
\usepackage{amsmath}

\begin{document}

\twocolumn[ %% activate for two-column option

\title{Plasmonic Tamm states: second enhancement of light inside the plasmonic waveguide}
%% For REVTeX it is possible to automate superscript and e-mail callouts with the superscriptaddress option; see REVTeX4 documentation.

\author{Yinxiao Xiang, Pidong Wang, Wei Cai,$^{*}$ Xinzheng Zhang, Cuifeng Ying and Jingjun Xu$^{\ddag}$}        
\address{
The Key Laboratory of Weak Light Nonlinear Photonics, Ministry of Education, \\
TEDA Applied Physics School and School of Physics, Nankai University, Tianjin 300457, China \\
$^*$weicai@nankai.edu.cn  \\
$^{\ddag}$jjxu@nankai.edu.cn}
\begin{abstract}
A type of Tamm states inside metal-insulator-metal (MIM) waveguides is proposed. An impedance based transfer matrix method is adopted to study and optimize it. With the participation of the plasmonic Tamm states, fields could be enhanced twice: the first is due to the coupling between a normal waveguide and a nanoscaled plasmonic waveguide and the second is due to the strong localization and field enhancement of Tamm states. As shown in our 2D coupling configuration, $|E|^2$ is enhanced up to 1050 times when 1550~nm light is coupled from an 300~nm Si slab waveguide into an 40~nm MIM waveguide.
\end{abstract}

\ocis{240.6680, 250.4480, 250.5403.}

 ] %% activate for two-column option
\noindent The delivery and concentration of optical radiation energy are essential for nano-optics such as plasmonic chips~\cite{ozbay2006plasmonics}, nanolithography~\cite{srituravanich2004plasmonic} and nanosensing~\cite{ringler2008shaping}. It is common to focus light by gradually decreasing a waveguide's cross section. Surface plasmon polaritons (SPPs), excitons propagating at the interface between a metal and dielectric, have been proposed to overcome the diffraction limit and concentrate light efficiently~\cite{Maier.07, Barnes.03}. SPPs based focusing configurations such as dielectric wedges~\cite{verhagen2010plasmonic}, tapered metal lateral waveguides~\cite{verhagen2008nanofocusing}/tips~\cite{stockman2004nanofocusing, ropers2007grating, issa2007optical}/grooves~\cite{volkov2009nanofocusing, choi2009compressing}/pyramids~\cite{lindquist2010three} and tapered metal-insulator-metal (MIM) waveguides~\cite{feng2007plasmon, choo2012nanofocusing, verhagen2009nanowire}/closed gaps~\cite{sondergaard2009resonant}/dimple lens~\cite{vedantam2009plasmonic} have been proposed theoretically or demonstrated experimentally. However, maximum enhancements among these proposals are always accompanied by a requirement of a sub-10~nm taper width, which is high-cost and difficult to fabricate. Ohm damping grows as the waveguide width decreases. Meanwhile, nonlocal effects will also appear in this region and lower the enhancement~\cite{garcia2008nonlocal, mcmahon2009nonlocal}. Due to above reasons the deviations between the predicted maximal values and actual results are always large. Is it possible to acquire considerable field enhancement without such a sharp taper end?

Optical Tamm states (OTSs, also known as Tamm plasmons), was recently discovered at the interface between a metal film and an 1D photonic crystal (PC)~\cite{kaliteevski2007tamm, sasin2008tamm}. Unlike the only TM-polarized SPPs, OTSs can be excited directly by normal incidence for both TM and TE-polarized waves. The light frequency lies in the band gap of the PC. Fields are confined and amplified at the metal/PC interface due to the interference of light reflected from both sides. Owing to the strong field enhancement, applications such as lasers~\cite{symonds2013confined, gazzano2011evidence}, all-optical switches~\cite{zhang2013all}, solar cells~\cite{zhang2012optical}, Faraday rotation~\cite{goto2009tailoring, da2013monolayer} and electromagnetically induced transparency (EIT)~\cite{lu2013optical} have been proposed.

In this letter, we introduce and demonstrate a type of Tamm states in plasmonic metal-insulator-metal (MIM) waveguides. In analogy with OTSs in an optical system, such states exist in a plasmonic system, as which we call plasmonic Tamm states (PTSs). We form Bragg reflectors (BR) by periodically change the dielectric materials in MIM~\cite{hosseini2006low}. Only if the phase matching condition $r_{\rm{left}}r_{\rm{right}}=1$ is fulfilled, PTSs can be generated in accompany with fields enhancement at the interface between the MIM BR and the MIM end~\cite{kaliteevski2007tamm}. Hence, fields can be enlarged twice during the focusing process: the first enhancement is due to the coupling between a normal dielectric waveguide and a nanoscaled MIM waveguide and the second enhancement is due to the strong localization and amplification of PTSs.
\begin{figure}[htb]
\centerline{\includegraphics[width=8.3cm]{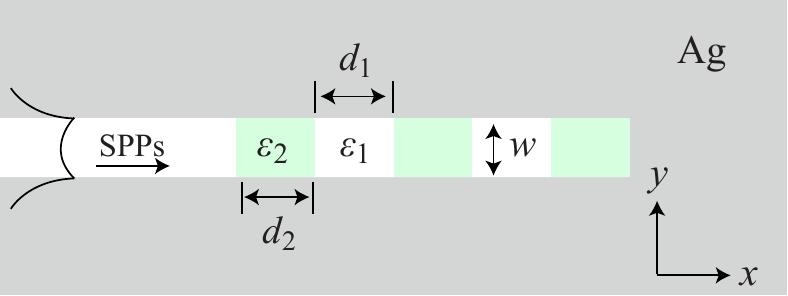}}
\caption{Sketch of the plasmonic Tamm states generation configuration. In order to fulfill the phase matching condition, $\varepsilon_2$ should be larger than $\varepsilon_1$.}\label{f1}
\end{figure}

\begin{figure}[htb]
\centerline{\includegraphics[width=8.3cm]{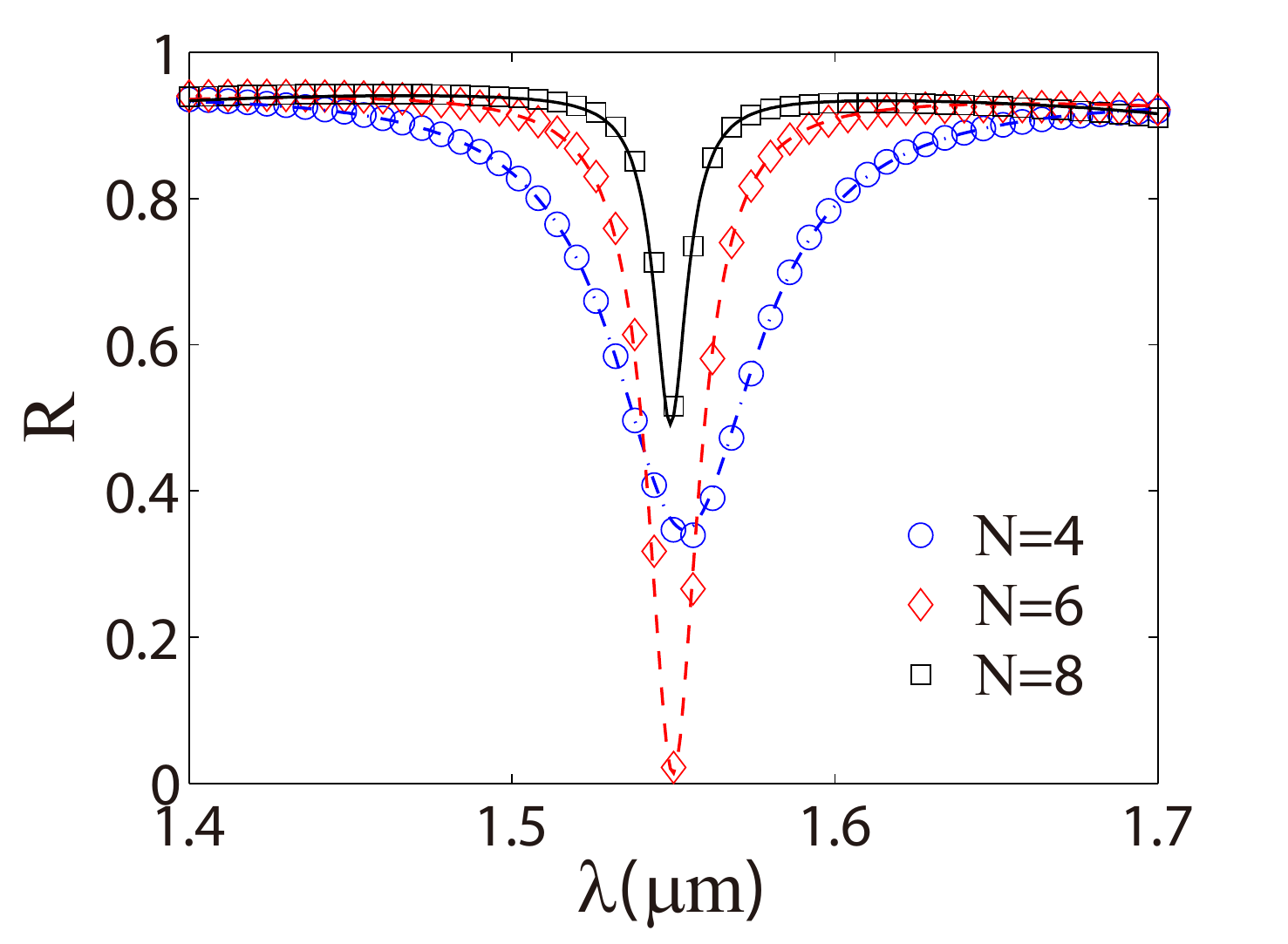}}
\caption{Comparison of reflection spectra calculated by TMM ($N$=4 (dash-dot), 6 (dashed) and 8 (solid)) and FEM (N=4 (open circle), 6 (open diamond) and 8 (open rectangle)). The reflection reaches a minimum at around 1.4\% for $N$=6.}\label{f2}
\end{figure}

The scheme of the PTSs generation configuration is illustrated in Fig.~\ref{f1}. SPPs are excited at the left side. The operation wavelength is 1550~nm. The dielectric core has a width of $w=40$~nm. In this regime, the odd TM mode ($E_x$ is odd, $E_y$ and $H_z$ are even) is the only mode of this MIM structure with a dispersion relation~\cite{Maier.07}
\begin{eqnarray}
\label{dispersion}
-\tanh\left(\frac{k_{d}w}{2}\right)=\frac{k_{m}\varepsilon_{d}}{k_{d}\varepsilon_{m}}
\end{eqnarray}
where $\varepsilon_{m}$ and $\varepsilon_{d}$ are the dielectric constants of the metal and the dielectric, respectively. $k_{m}$ and $k_{d}$ denote the respective transverse propagation constants. The metal is silver, and the dielectrics are set as $\varepsilon_1=1$ and $\varepsilon_2=2.25$. When PTSs are excited, fields are enhanced, trapped and also significantly Ohm damped at the interface between the MIM BR and the MIM end. Therefore, a minimum in reflection may reveal a maximum in enhancement. Since the reflection and transmission of MIM structures could be characterized by waveguide impedances, we can use an impedance based transfer matrix method (TMM)~\cite{hosseini2008modeling} to find the optimal parameters of the BR, which are $d_1=259$~nm, $d_2=171$~nm. Reflection spectra for MIM BR with periods $N=4$ (dash-dot), 6 (dashed) and 8 (solid) are plotted in Fig.~\ref{f2}. Our results are confirmed by the finite element method (FEM) based commercial software Comsol Multiphysics. The FWHM decreases as $N$ increases. The reflection reaches a minimum at around 1.4\% with a FWHM of 18~nm for $N$=6, which also enable it to be a good narrow-band absorber~\cite{gong2011perfect}.

\begin{figure}[htb]
\centerline{\includegraphics[width=8.3cm]{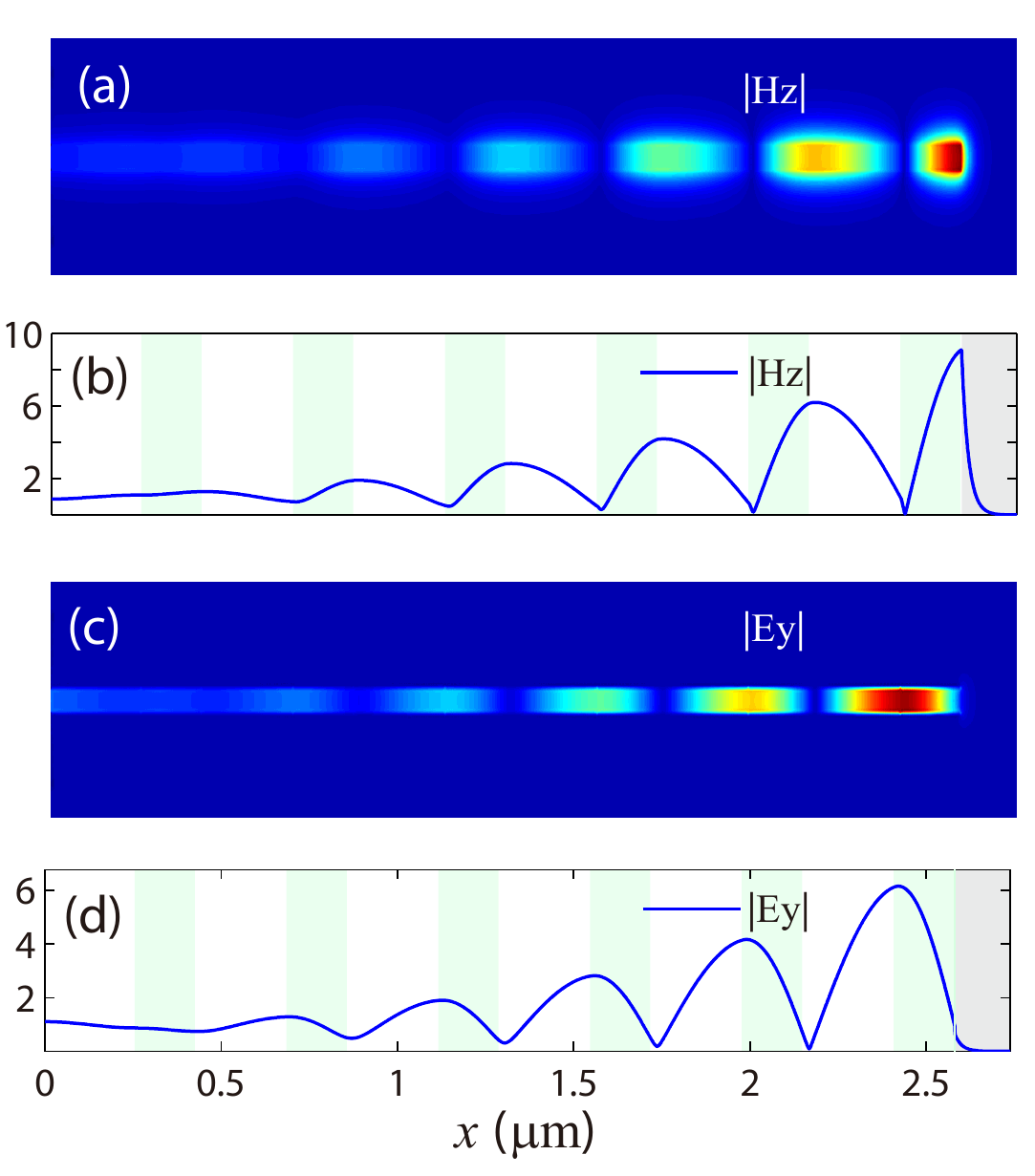}}
\caption{Field amplitude distribution in side the MIM structure. Both magnetic and electric fields are normalized by the incoming ones in order to show the enlargement factor directly. The maximal enlargement factor of $H_z$ is 9.1~times and $E_y$ 6.2~times.}\label{f3}
\end{figure}

In Fig.~\ref{f3}~(a) and (c), we plot the field amplitude distribution $|H_z|$ and $|E_y|$ for $N$=6. Fields are trapped inside the MIM structure. They reach to maxima at around the interface between the MIM BR and the MIM end. Cutlines along $y=0$ are plotted in Fig.~\ref{f3}~(b) and (d). Both fields are normalized by the incoming ones in order to show the enhancement directly. Field amplitudes oscillate in the MIM BR and decay exponentially in the MIM end. The maximal $|H_z|$ occurs at the interface between the BR and the MIM end by an enlargement factor of 9.1. The maximal $|E_y|$, which occurs at the first $\varepsilon_1$/$\varepsilon_2$ interface near the MIM end, is 6.1 times larger than the incidence. It should be noted that positions for each $|E_y|$ local maxima is also related to each $|H_z|$ local minima, and vice versa.

\begin{figure}[htb]
\centerline{\includegraphics[width=8.5cm]{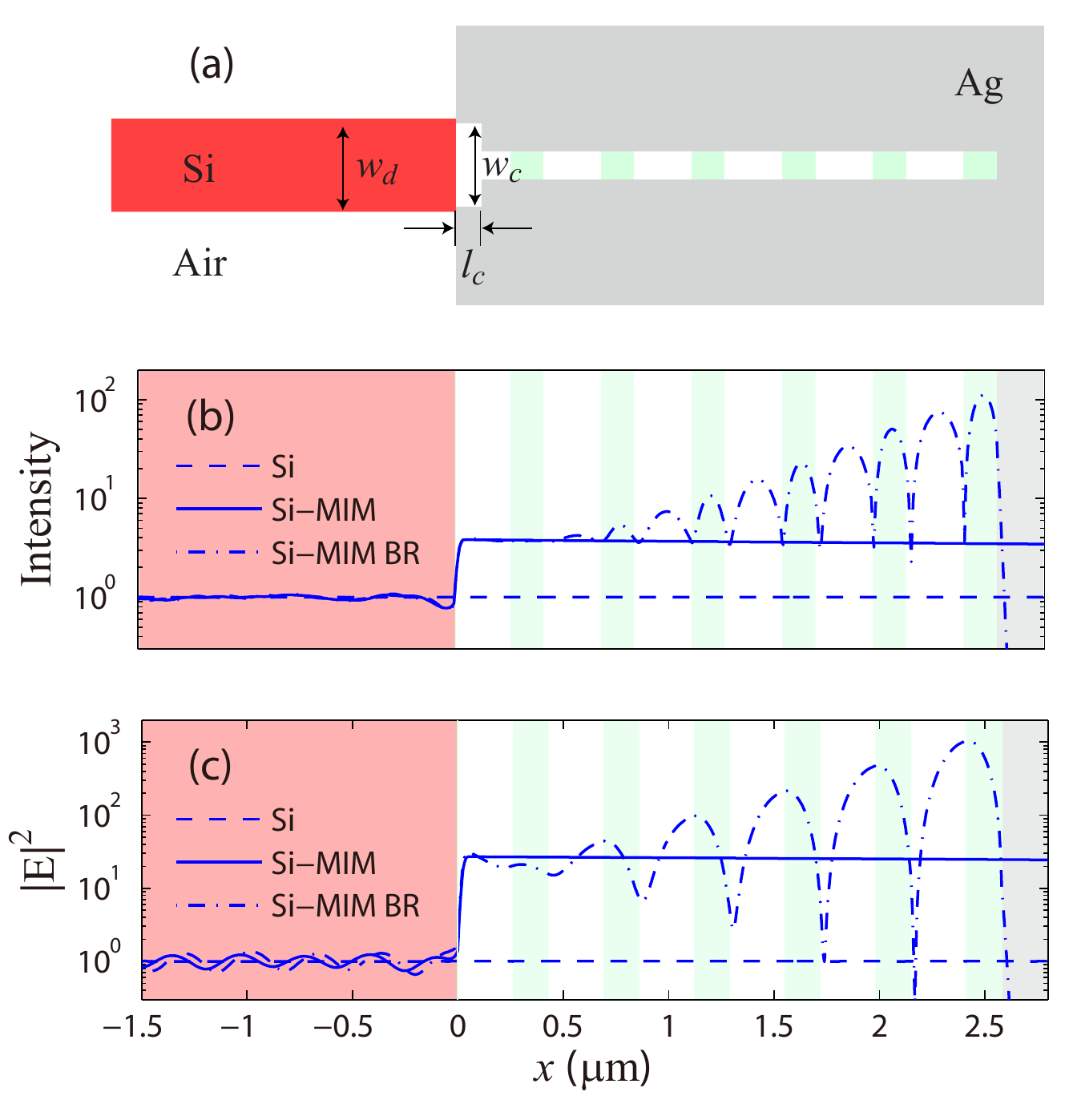}}
\caption{(a) The 2D coupling configuration. (b) Field intensity distribution and enhancement. The maximal enlargement factor is 110-fold. (c) The $|E|^2$ distribution and enhancement. The maximal enhancement is 1050-fold.}\label{f4}
\end{figure}

Finally, we turn to the 2D coupling configuration as depicted in Fig.~\ref{f4}~(a). Parameters for the MIM structure are unchanged. Light is propagating along an Si slab waveguide with a width $w_d$=300~nm. It is coupled into the MIM waveguide by an air gap coupler with $w_c$=292~nm and $l_c$=28~nm~\cite{Wahsheh.09}. The coupling efficiency is around 90\%. As shown in Fig.~\ref{f4} (b), intensities for light propagating in a slab waveguide alone (dashed), a slab waveguide meets an MIM waveguide (solid) and a slab waveguide meets an MIM BR (dash-dot) are plotted. Data are normalized by the incoming intensity. The first field enlargement is due to the dimension confinement as shown by the solid line. The enlargement factor is 3.8. Then, the intensity oscillates and be further amplified inside the MIM BR when PTSs are excited. The maximal enhancement is 110-fold. As there is an alternative $|H_z|/|E_y|$ local minimum at each $\varepsilon_1$/$\varepsilon_2$ interface, the local minimal intensities occurs at these positions as well. We also plot the $|E|^2$ distribution because optical phenomenons such as nonlinearities and surface enhanced Raman scattering are related to it rather than the intensity. As shown in Fig.~\ref{f4}~(c), the first enhancement factor by the dimension confinement is 27-fold (solid). After the second enhancement by PTSs, the maximal enhancement is 1050-fold. The local maximum of $|E|^2$ occurs each period.

In summary, we have proposed and studied PTSs inside MIM plasmonic waveguides. By using the impedance based TMM, MIM BR is optimized efficiently. We also show an 2D coupling configuration in which light is coupled from an 300~nm slab waveguide into an 40~nm MIM waveguide by an air-gap coupler. With the participation of PTSs, $|E|^2$ is amplified by as large as 1050~times. We strongly believe that PTSs will find great applications in intensity-dependent nano-optics.

This work was financially supported by the National Basic Research Program of China (2010CB934101, 2013CB328702), the National Natural Science Foundation of China (11004112), the 111 Project (B07013), International S\&T Cooperation Program of China (2011DFA52870), and Oversea Famous Teacher Project (MS2010NKDX023), International cooperation program of Tianjin (11ZCGHHZ01000).

%\bibliography{D:/reference/AIP/ref}   %>>>> bibliography data in report.bib\
%\bibliographystyle{ol}   %>>>> makes bibtex use spiebib.bst

%\pagebreak
%\bibliography{D:/reference/AIP/ref}   %>>>> bibliography data in report.bib\
%\bibliographystyle{osajnl}   %>>>> makes bibtex use spiebib.bst

\end{document}